\documentclass[prd,11pt,preprint,nofootinbib,showkeys]{revtex4-1}
\usepackage{graphics}%
\usepackage[mathcal]{euscript}
\usepackage[latin1]{inputenc}
\usepackage{latexsym}
\usepackage{hyperref}
\usepackage{amsmath}    %
\usepackage{graphicx}   %
\usepackage{verbatim}  
\usepackage{color}      
\usepackage{subfigure}  
\usepackage{hyperref}   
\usepackage{bm}
\usepackage{graphicx}
\usepackage{amssymb}
\usepackage{bbm}
\usepackage{float}
\usepackage{slashed}
\usepackage{amsfonts}
\usepackage{euscript}
\usepackage{mathrsfs} 
\usepackage{bbding}
\usepackage{pifont}
\usepackage{cancel}
\usepackage{multirow}
\usepackage{soul}
\usepackage{physics}

\usepackage[usenames,dvipsnames]{xcolor}
\hypersetup{colorlinks=true, citecolor=Aquamarine, linkcolor=Aquamarine,urlcolor=Aquamarine}

\makeindex

\begin{document}

\title{Toward a mechanism for the emergence of gravity}

\author{Carlos Barcel\'o}
\email{carlos@iaa.es}
\affiliation{Instituto de Astrof\'{\i}sica de Andaluc\'{\i}a (IAA-CSIC), Glorieta de la Astronom\'{\i}a, 18008 Granada, Spain}
\author{Ra\'ul Carballo-Rubio}
\email{raul.carballorubio@ucf.edu}
\affiliation{Florida Space Institute, University of Central Florida, 12354 Research Parkway, Partnership 1, Orlando, Florida 32826, USA}
\author{Luis J. Garay}
\email{luisj.garay@ucm.es}
\affiliation{Departamento de F\'{\i}sica Te\'orica and IPARCOS, Universidad Complutense de Madrid, 28040 Madrid, Spain}
\affiliation{Instituto de Estructura de la Materia (IEM-CSIC), Serrano 121, 28006 Madrid, Spain}
\author{Gerardo Garc\'ia-Moreno}
\email{gerargar@ucm.es}
\affiliation{Departamento de F\'{\i}sica Te\'orica and IPARCOS, Universidad Complutense de Madrid, 28040 Madrid, Spain}
\affiliation{Instituto de Astrof\'{\i}sica de Andaluc\'{\i}a (IAA-CSIC), Glorieta de la Astronom\'{\i}a, 18008 Granada, Spain}

\begin{abstract}
One of the main problems that emergent-gravity approaches face is explaining how a system that does not contain gauge symmetries \emph{ab initio} might develop them effectively in some regime. We review a mechanism introduced by some of the authors for the emergence of gauge symmetries in [JHEP 10 (2016) 084] and discuss how it works for interacting Lorentz-invariant vector field theories as a warm-up exercise for the more convoluted problem of gravity. Then, we apply this mechanism to the emergence of linear diffeomorphisms for the most general Lorentz-invariant linear theory of a two-index symmetric tensor field, which constitutes a generalization of the Fierz-Pauli theory describing linearized gravity. Finally we discuss two results, the well-known Weinberg-Witten theorem and a more recent theorem by Marolf, that are often invoked as no-go theorems for emergent gravity. Our analysis illustrates that, although these results pinpoint some of the particularities of gravity with respect to other gauge theories, they do not constitute an impediment for the emergent gravity program if gauge symmetries (diffeomorphisms) are emergent in the sense discussed in this paper.

\end{abstract}


\maketitle

{

\hypersetup{citecolor=black, linkcolor=black,urlcolor=black}

\tableofcontents

}

\section{Introduction}
\label{Introduction}
The pursuit of a theory able to combine the principles of general relativity and quantum mechanics is one of the strongest driving forces in modern fundamental physics. Almost a century of active research on this problem has resulted in many insights~\cite{Carlip2001}, although a full comprehension of how both sets of principles might intertwine is still lacking. Approaches toward finding a theory of quantum gravity can be divided in emergent and non-emergent approaches. The first category corresponds to approaches in which the fundamental degrees of freedom are not taken to be the spacetime geometry itself, but other sort of microscopic degrees of freedom. The geometric structure characteristic of general relativity is assumed to be recovered in a suitable regime, for instance a weakly-coupled regime or a low-energy regime. String theory~\cite{Polchinski1998a,Polchinski1998b} would fall in this category, being a specific proposal for these microscopic degrees of freedom. More generally, one can conceive systems having another substratum for the emergence, for instance condensed-matter systems~\cite{Volovik2008,Volovik2009,Barcelo2005,Barcelo2014a}. The second category corresponds to theories in which the geometric structure of general relativity is taken as fundamental. Different approaches in this category are set apart by the different quantization schemes applied to this structure. The most widely studied theory within this category is loop quantum gravity~\cite{Thiemann2007}. The distinction between the two categories above is nevertheless not a sharp one. In fact, the notion of emergence is inevitably tied up to every quantum gravitational theory, since one can show on general grounds that there are no local observables in the Dirac sense~\cite{Dirac1964} (that is, diffeomorphism invariant) for any quantum gravity theory~\cite{Carlip2013}. 

This work is devoted to analyzing a specific kind of systems within the emergent category. The motivation for this study comes from the observation that it is quite common to find analogues of gravitational fields in many physical systems, specially in condensed-matter systems~\cite{Barcelo2005}. However, this analogy just holds at a kinematical level, meaning that the effective geometry emerging in those systems does not display the dynamical features characteristic of general relativity. Thus, a natural question that arises is whether these analogue systems can be designed to develop the correct dynamics. This is the notion of emergent gravity that we explore in this work. Our goal is finding the conditions that make such analogue systems develop a fully relativistic dynamics in some regime. 

The emergence of analogue gravity usually comes associated with a low-energy regime in which the excitations can be typically described by a collection of weakly-coupled relativistic fields. They describe the collective excitations of this system~\cite{Barcelo2014a}. In condensed-matter systems, this low-energy regime is usually characterized by the presence of Fermi points~\cite{Volovik2008,Volovik2009}. Around these Fermi points, we know the field content relevant for describing the low-energy regime, and we know that the fields are weakly coupled. This low-energy Lorentz invariance becomes less exact as we increase the energy. It typically manifests itself in modified dispersion relations of the form
\begin{equation}
    E^2 = k^2  + \sum_n \xi_n \frac{k^{2n + 2}}{E_L^{2n}},
\end{equation}
for instance, but also on the very dissolution of the effective description. The dispersion relations by themselves offer a problem from an effective field theory rationale since they appear to generate a fine-tuning problem (see ~\cite{Collins2004}). Here we do not delve into this problem, which we think arises because of the tacit assumption that the effective field description survives far in the dispersive regime, and just assume the existence of a low-energy Lorentzian regime. This ensures that we can apply the rules of effective field theory and write down the most general effective action compatible with the symmetries of the system in order to obtain a good description of the phenomena of interest~\cite{Polchinski1992,Manohar2018}. However, these arguments cannot guarantee that a vector field $A^{\mu}$ or a symmetric tensor $h^{\mu \nu}$ will develop gauge symmetries. It is well known that gauge symmetries prevent the propagation of ghost states and hence eliminate the potential instabilities of such theories. The absence of gauge symmetries would leave us with a potentially unstable theory. Hence, the low-energy description needs to be endowed with a mechanism that allows the effective decoupling of the dangerous degrees of freedom. We discuss here a specific mechanism for this decoupling, first introduced by some of us in~\cite{Barcelo2016} for an extended version of (quantum) electrodynamics.

Here is a brief outline of this work. In Sec.~\ref{Mechanism} we review on general grounds the mechanism for the emergence of gauge symmetries in systems displaying no gauge symmetries~\emph{ab initio}. Section~\ref{Examples} is devoted to discuss some examples. In Subsec.~\ref{Example1} we review how this mechanism works for an extended version of linearized Yang-Mills theory and how a non-linear completion of such theory can be obtained through a bootstrapping procedure, which was studied in detail in~\cite{Barcelo2021} where further details can be found. In Subsec.~\ref{Example2}, we consider the most general linear theory that can be written down for a symmetric tensor field $h^{\mu \nu}$. We study the mechanism for the emergence of gauge symmetries introduced and find the emergence of the closest gauge symmetries to diffeomorphisms and WTDiff within this framework. We discuss the relation of this result with the emergence of linearized gravity and discuss ongoing work on the possible non-linear completions of the theory. In Sec.~\ref{No-gos} we revisit some theorems, namely the Weinberg-Witten theorem in Subsec.~\ref{WW} and Marolf's theorem~\ref{Marolf}, that are often invoked as no-go theorems for emergent gravity and discuss their limitations and interplay with our results. We conclude in Sec.~\ref{Conclusions}. Furthermore, we have added Appendix~\Ref{sec:appendix} where we discuss some specific cases associated with particular choices of parameters for the $h^{\mu \nu}$-theory from Subsection~\ref{Example2} that make the discussion of the emergence of gauge symmetries slightly different.

\textit{Notation and conventions.} We work in four-dimensional spacetime and we use the signature $(-,+,+,+)$. The symbol $\nabla$ represents the covariant derivative compatible with the flat metric of Minkowski spacetime $\eta_{\mu \nu}$. We use greek $(\mu, \nu...)$ indices for the spacetime indices and latin indices $(a,b...)$ for the internal indices of vector fields.

\section{A mechanism for the emergence of gauge symmetries}
\label{Mechanism}

The first step in the logic of our construction is to consider a non-relativistic condensed-matter-like theory that in some regime can be accurately described in terms of a set of Lorentz-invariant and weakly coupled fields over a flat background. The emergence of Lorentz invariance in this kind of systems is typically associated with the presence of Fermi-like points in the theory~\cite{Volovik2008,Volovik2009,Barcelo2014a}. Here, we take for granted this first step and analyze under which conditions this effective field theoretic description develop emergent gauge symmetries.

We will work in configuration space to avoid introducing Dirac's machinery for the classification of constrained systems~\cite{Dirac1964}. Thus, we need a characterization of gauge symmetries within this formulation. Such characterization is offered by Noether's theorem~\cite{Henneaux1992}. Generally, given a symmetry we can find a conserved current associated with it. If the symmetry is a gauge symmetry, the Noether charges associated with the corresponding current are trivial. Hence, for a gauge symmetry the current can always be written as 
\begin{equation}
    J^{\mu} = W^{\mu} + S^{\mu},
    \label{gauge_current}
\end{equation}
with $W^{\mu}$ being a term that is zero on shell and $S^{\mu} = \nabla_{\nu} N^{\nu \mu}$ a superpotential, i.e. an identically conserved term. Once the theory is provided with suitable fall-off conditions, the contribution of the superpotential term to the charges identically vanishes, as it can be seen by applying Gauss theorem~\cite{Julia1998}. 

This characterization of gauge symmetries is the starting point for our discussion of emergent gauge symmetries. Let us consider a field theory characterized by a set of fields, which we call $\Phi$, with a well-posed initial value problem. For the sake of simplicity, we assume that such system is free of gauge symmetries from the beginning. There always exists a set of symmetries whose Noether charges parametrize the space of initial conditions~\cite{Aldaya2016}, and hence the whole space of solutions which we call $\mathcal{S}$. These symmetries are in general complicated symmetries that cannot be found explicitly; they can only be found for integrable systems. Knowing that they exist is enough for our purposes. 

Now let us consider a set of constraints defined by the condition $\Psi = 0$. Here, $\Psi$ represents a set of equations that combine the fields $\Phi$ and their derivatives. Let us represent the subspace defined by this constraint as $\mathcal{U} \subset \mathcal{S}$. We demand such space to be non-trivial in the sense of having more than one element. This means that there exist solutions for the constraint $\Psi = 0$ apart from the trivial one in which all the fields vanish, $\Phi = 0$.

Let us denote $\mathcal{Q}$ the complete set of charges that parametrizes the space of solutions $\mathcal{S}$. It is convenient to choose a parametrization of the space of charges, i.e. a set of coordinates in charge space, such that the subspace $\mathcal{U}$ is characterized by the condition that some of the charges vanish. In other words, we choose a subset of charges $\mathcal{Q}^{\perp}_{\Psi} \subset \mathcal{Q}$ such that $\Psi = 0 $ if and only if $\mathcal{Q}^{\perp}_{\Psi}  = 0$. Furthermore, it is useful for our purposes to choose the remaining coordinates, which we denote as $\mathcal{Q}^{\parallel}_{\Psi}$, in such a way that they leave the subspace $\mathcal{U}$ untouched. This means that the symmetries associated with the charges $\mathcal{Q}^{\parallel}_{\Psi}$ can be defined within $\mathcal{U}$ since they preserve the condition $\Psi = 0$. Now we have two possibilities depending on the values of the charges $\mathcal{Q}^{\parallel}_{\Psi}$ within the subspace $\mathcal{U}$. 

\textbf{Non-emergence of gauge symmetry:} This corresponds to the standard situation in which the charges $\mathcal{Q}^{\parallel}_{\Psi}$ parametrize the space $\mathcal{U}$ without redundancies. In this case, we will have that the projection onto the subspace $\mathcal{U}$ simply eliminates the degrees of freedom directly encoded in the conditions $\Psi = 0$.

\begin{figure}[H]
\begin{center}
\includegraphics[width=9.0 cm]{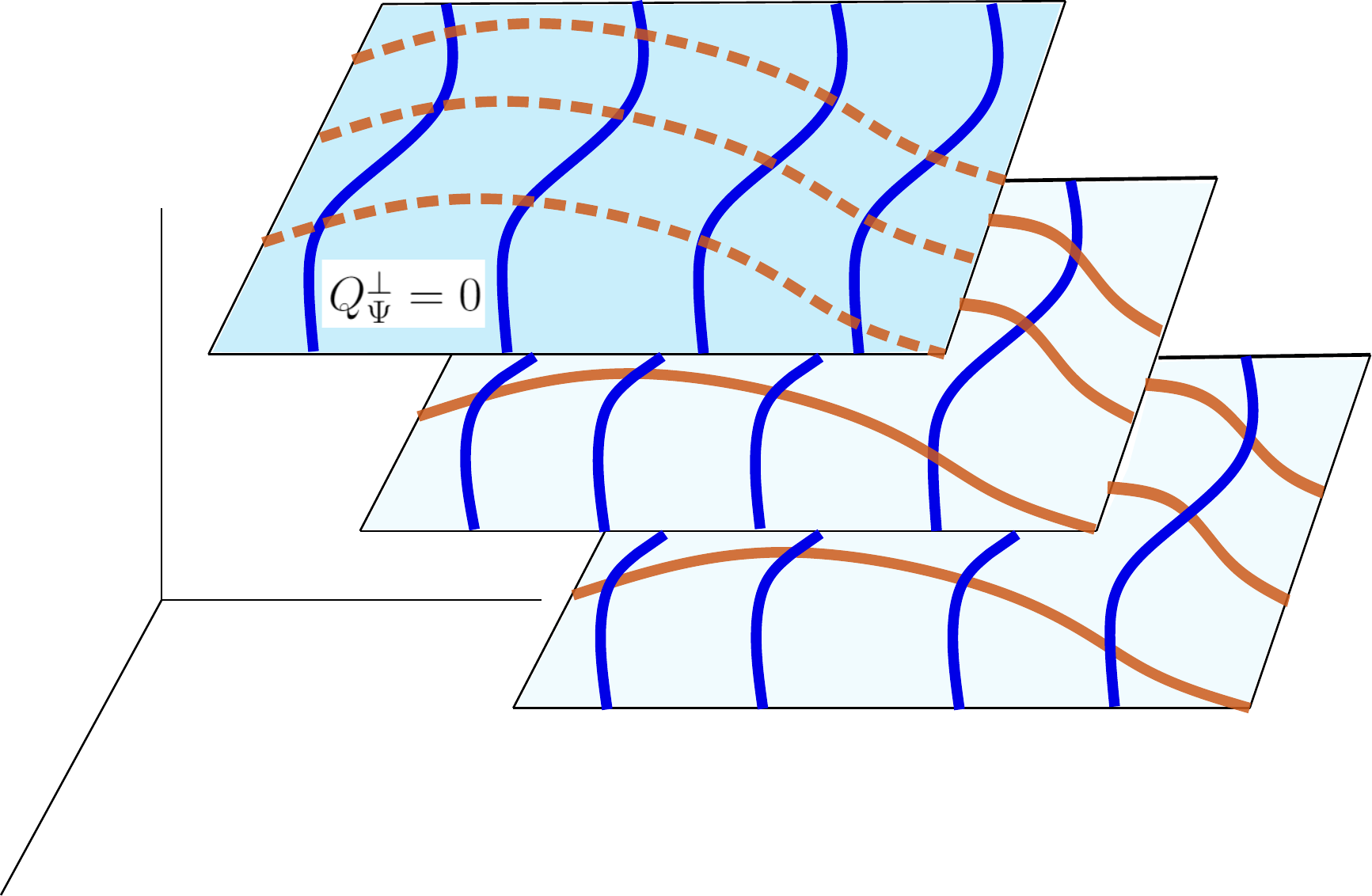}
\caption{We represent the space of solutions $\mathcal{S}$ and the set of surfaces characterized by constant values of $\mathcal{Q}^{\perp}_{\Psi}$. Among them, there is a privileged one represented in darker blue characterized by $\mathcal{Q}^{\perp}_{\Psi} = 0$, i.e. it represents the subspace $\mathcal{U}$ from the discussion above. The solid blue and brown curves represent curves of constant $Q_{\Psi}^{\parallel}$ within the surfaces of $\mathcal{Q}^{\perp}_{\Psi} = \textrm{constant}$. The dashed brown curves within the subspace $\mathcal{U}$ represent the emergent gauge orbits, i.e. they connect points that cannot be distinguished by performing operations within $\mathcal{U}$ and are characterized by the same vanishing value of some of the charges $Q_{\Psi}^{\parallel}$ in $\mathcal{U}$.}
\end{center}
\label{Fig:emergence}
\end{figure}

\textbf{Emergence of gauge symmetry:} This case corresponds to the situation in which some of the charges $\mathcal{Q}_{\Psi}^{\parallel}$ vanish when we restrict to $\mathcal{U}$. In this case, not all the solutions in $\mathcal{U}$ can be distinguished by just using operations within $\mathcal{U}$. Hence, the solutions related by a finite transformation associated with the charges $\mathcal{Q}_{\Psi}^{\parallel}$ can be interpreted as describing a single physical configuration. A pictorical representation of this situation can be found in Fig.~\Ref{Fig:emergence}. 

Interpreting these configurations as a single physical configuration means that these finite transformations would correspond to emergent gauge symmetries. Furthermore, we would interpret the orbits generated by these transformations as emergent gauge orbits. This translates into the elimination of entire dynamical degrees of freedom, each corresponding to a pair of initial conditions or charges that parametrize $\mathcal{S}$. Notice that this mechanism relies on the naturalness of the condition $\Psi = 0$ in some regime of the theory, e.g. a low-energy regime or a weakly-coupled one. We will come back to this point when considering specific examples below.

This mechanism has been implemented successfully for Maxwell~\cite{Barcelo2016} and linearized Yang-Mills theories~\cite{Barcelo2021}. To extend this mechanism to non-linear theories, in~\cite{Barcelo2021} we demanded the existence of a bootstrapping procedure that determines the self-coupling of the linear theory, thus connecting smoothly the linear and non-linear theories. In this paper we start exploring the application of this mechanism to gravity.

\section{Two examples: Yang-Mills theories and linearized gravity}
\label{Examples}

\subsection{Yang-Mills theories}
\label{Example1}
Let us present here an extended version of Yang-Mills theory that will allow us to illustrate the emergence of gauge symmetries. Our starting point is the most general quadratic and Lorentz-invariant lagrangian giving rise to second-order differential equations that can be written for a set of $N$ vector fields $A^{a}_{\mu}$, with the latin indices ($a,b,c...)$ running from $1$ to $N$:
\begin{equation}
    S_2 = \int d^4 x \sqrt{-\eta} \left[ - \frac{1}{4}F^{a}_{\mu \nu} F_{a}^{\mu \nu} + \frac{1}{2} \xi_{a b} ( \nabla_{\mu} A^{a \mu} ) ( \nabla_{\nu} A^{b \nu} ) - \frac{1}{2} M_{a b} A^{a \mu} A^{b}_{\mu} + A_{a \mu} j^{a \mu} \right],
    \label{action_principle}
\end{equation}
up to a boundary term which we skip here. We have introduced the notation ${F^{a}_{\mu \nu} = 2 \nabla_{[\mu} A^{a}_{\nu]}}$, where $\nabla$ is the affine connection compatible with $\eta_{\mu \nu}$, and we have introduced the non-degenerate matrix $\xi_{a b}$ and the mass matrix $M_{ab}$. The case which is specially interesting for us is the massless case $M_{ab} = 0$, since it is a generic feature that field theories emerging around Fermi points in condensed-matter systems give rise to massless excitations~\cite{Volovik2009}. Masses can be acquired in a second step through a Higgs-like mechanism. However, for the moment, we will keep the $M_{a b}$ term since it is instructive to study the massive and massless cases as representatives of the two possible situations described in the previous section regarding the mechanism for the emergence of gauge symmetries. The equations of motion can be derived straightforwardly from the action~\eqref{action_principle} by taking a simple functional derivative,
\begin{equation}
    \nabla_{\mu} F_{b}^{\mu \nu} - \xi_{ab} \nabla^{\nu} \nabla_{\mu} A^{a \mu} - M_{ab} A^{a \nu} = j_{b}^{\nu}.
    \label{eoms_ym}
\end{equation}
It is possible to identify a set of symmetries for this theory associated with the following transformations: 
\begin{equation}
    A^{a}_{\mu} \rightarrow A^{a}_{\mu} + \nabla_{\mu} \chi^a, \qquad j^{a \mu} \rightarrow j^{a \mu},
    \label{gauge_transf_ym1}
\end{equation}
where $\chi^a$ are functions that need to obey 
\begin{equation}
    \left( \xi_{a b} \Box + M_{a b} \right) \chi^a = 0.
    \label{gauge_transf_ym2}
\end{equation}
The currents associated with these symmetries, once evaluated on shell, can be written as follows~\cite{Barcelo2021}:
\begin{equation}
    J^{\mu} \vert_{\textrm{on shell}} = - \nabla_{\nu} \left( F^{a \mu \nu} \chi_a \right) + \xi_{a b} \left( \nabla_{\nu} A^{a \nu} \nabla^{\mu} \chi^{b} - \chi^{b} \nabla^{\mu} \nabla_{\nu} A^{\nu a} \right)
\end{equation}
Notice that the first part of this equation is a superpotential and hence it does not contribute to the charges. Furthermore, the second term vanishes if we restrict ourselves to the subspace $\mathcal{U}$ of solutions of Eq.~\eqref{eoms_ym} satisfying $\nabla_{\mu} A^{a \mu} = 0$. Hence, projecting onto the subspace $\mathcal{U}$ makes the current acquire the form of a gauge current encapsulated in Eq.~\eqref{gauge_current}. Notice that although Eq.~\eqref{current_on_shell} is associated with a transformation whose generators $\chi^a$ are spacetime-dependent local functions, it does not constitute a gauge transformation \emph{a priori}. Gauge transformations are associated with local transformations whose generators are arbitrary functions of spacetime. However, in our case they are constrained to obey Eq.~\eqref{gauge_transf_ym2}. Thus, the transformation we are discussing does not constitute a gauge transformation from the beginning although it is generated by local functions. A further discussion of this point with a concrete example can be found in Secs. 2.1 and 2.2 from~\cite{Barcelo2016}.

The projection introduced above is quite natural once we consider the dynamical equations that the scalar fields $\varphi^a = \nabla_{\mu} A^{a \mu}$ obey, which can be obtained by taking the divergence in Eq.~\eqref{eoms_ym}:
\begin{equation}
    \left[  \Box \delta^{c}_{\ a}+ \left(\xi^{-1} \right)^{cb} M_{a b} \right] \varphi^{a} = 0.
\end{equation}
We have a set of sourceless Klein-Gordon equations for these scalars $\varphi^a$, and hence it is quite natural to restrict ourselves to the subspace $\mathcal{U}$ defined by $\varphi^a = 0$, since there are no Lorentz-invariant sources that might generate excitations of these scalars. From this point onwards, there are two possible situations which we discuss below. 

\textbf{Massive case $M_{ab} \neq 0$.} In this case no gauge symmetries emerge, since the putative emergent gauge symmetries from Eqs.~\eqref{gauge_transf_ym1} and~\eqref{gauge_transf_ym2} do not leave $\mathcal{U}$ invariant. Actually, they translate into the following map for $\varphi^a$
\begin{equation}
    \varphi^a \rightarrow \varphi^a - \left( \xi^{-1} \right)^{a b} M_{b c} \chi^{c},
    \label{U_mapping}
\end{equation}
which takes $\varphi^a = 0$ to a non-vanishing value for $\varphi^a \neq 0$.

\textbf{Massless case $M_{ab} = 0$.} In this case gauge symmetries emerge naturally, since the transformations from Eqs.~\eqref{gauge_transf_ym1} and~\eqref{gauge_transf_ym2} become gauge symmetries by leaving invariant the scalar fields $\varphi^a$ (note that the last term of~\eqref{U_mapping} now vanishes).

It is better to pause at this point to summarize what we have obtained so far and what are we pursuing for the non-linear theory. We began with the most general Lorentz-invariant vector field theory for a set of $N$ vector fields $A^{a}_{\mu}$. If these vector fields represent massless excitations, we have shown that the coupling to a conserved current leads to a set of sourceless Klein-Gordon equations for the scalar sector of the theory. Since no excitations can be produced within that subspace, we have removed the scalar sector by projecting onto the subspace $\varphi^a = 0$. The resulting effective field theory that we have found displays emergent gauge symmetries given by Eq.~\eqref{gauge_transf_ym1}, with $\chi^a$ satisfying $\Box \chi^a = 0$. Notice that these transformations are just the residual gauge symmetries of linearized Yang-Mills theory after the Lorenz gauge is chosen. 

The next natural step would be to study non-linear vector-field theories. Among the possible non-linear vector-field theories, Yang-Mills theories are special since their gauge symmetry avoids the propagation of ghosts when one linearizes them. Furthermore, they are such that they can be derived from the linear theory through a bootstrapping procedure in which one analyzes the self-coupling of the theory~\cite{Deser1970}. However, such procedure does not lead to a unique theory by itself due to the ambiguities inherent to the definition of Noether current~\cite{Ortin2010, Padmanabhan2004}, at least not with further specifications~\cite{Barcelo2014}. The combination of the mechanism introduced here for the emergence of gauge symmetries with the bootstrapping procedure allowed us to make an incursion on the non-linear vector-field theories that are free from instabilities. We present here the main idea and refer the reader to~\cite{Barcelo2021} for further technical details. 

The idea of the bootstrapping procedure is the following. Consider a linear field theory displaying a symmetry and hence having a conserved current which we call $J_{1}^{a \mu}$. We assume that our theory to couples perturbatively to such current, in such a way that we recover a smooth limit for the free theory when we take the coupling constant to be zero. We call such coupling constant $g$. Thus, schematically we will have equations of motion of the form
\begin{equation}
    \frac{\delta S_2}{\delta A_{a \mu}} = g J_{1}^{a \mu} + \order{g^2}.
\end{equation}
To derive the right-hand side from an action principle, we would need to add a suitable interacting term $S_3$ to the action such that
\begin{equation}
    \frac{\delta S_3}{\delta A_{a \mu}} = J_{1}^{a \mu}.
\end{equation}
However, notice that $S_3$, which needs to be also symmetric under the transformations that we introduce for the bootstrapping procedure, would contribute to the conserved current due to Noether's theorem. Hence, we would obtain a contribution to the current of the form $g J_2^{a \mu}$ which we would want to derive from an action principle, from $S_4$. In this way, we would obtain an infinite set of constraints for the successive actions $S_n$ and $S_{n+1}$. More concretely, we would have
\begin{equation}
    \frac{\delta S_{n+1}}{\delta A_{a \mu}} = J_{n-1}^{a \mu},
\end{equation}
where $J_{n-1}^{a \mu}$ is obtained from $S_n$ through Noether's theorem. In this way, it is possible to obtain a potentially infinite set of equations which we would need to solve in order to find the non-linear theory described by the action 
\begin{equation}
    S = \sum_{n=2}^{\infty} g^{n-2} S_n.
\end{equation}

There are two points we want to stress. First, we need to remember that conserved currents are defined up to the addition of identically conserved quantities. Hence, at each order in the bootstrapping procedure we have ambiguities that need to be taken into account. For instance, in the gravitational context they are crucial to recover general relativity from Fierz-Pauli theory through this procedure~\cite{Butcher2009}. Second, at each order we need to ensure that the generated contribution to the action $S_{n+1}$ is still invariant under the symmetry transformations that were used to define the current $J_{n-1}^{a \mu}$. This imposes non-trivial constraints on the partial actions~\cite{Barcelo2021}. 

For the case of emergent Yang-Mills theories, we need to consider the following quadratic action as the starting point,
\begin{equation}
    S_2 = \int d^4x \sqrt{-\eta} \left[ -\frac{1}{4}F^{a}_{ \mu \nu} F_{a}^{\mu \nu} +\frac{\lambda}{2}   (\nabla_{\mu} A_{a}^{ \mu} ) (\nabla_{\nu} A^{a \nu}) \right],
\end{equation}
which is the action from Eq.~\eqref{action_principle} with $M_{a b} = 0$, as we demand for the emergence of gauge symmetries; we have omitted the material content which we will discuss later; and we have restricted ourselves to the case $\xi_{ab} = \lambda \delta_{ab}$ for the sake of simplicity. The symmetries that we can identify to begin the bootstrapping procedure are the following rigid transformations:
\begin{equation}\label{sym0}
    A^{a \mu} \rightarrow A^{a \mu} + f^{a b c} A^{\mu}_{b} \zeta_c,
\end{equation}
where $\zeta_c$ are arbitrary real constants and $f^{a b c}$ is a constant tensor antisymmetric in the first two indices, i.e. obeying $f^{a b c} = f^{[a b] c}$. 

It turns out that the bootstrapping procedure imposes two non-trivial constraints that must be fulfilled to be self-consistent~\cite{Barcelo2021}. On the one hand, the constants $f^{a b c}$ must be the structure constants of a semi-simple compact Lie Algebra for the symmetry in Eq.~\eqref{sym0} to be also a symmetry of $S_3$. On the other hand, we need to impose the constraint $\nabla_{\mu} A^{\mu}_a = 0$ to avoid breaking the bootstrapping procedure since, if we do not impose it, the equations of motion cannot be derived from an action principle. The resulting action from the bootstrapping procedure can be written as 
\begin{equation}
    S = \int d^4x \sqrt{-\eta} \left[ -\frac{1}{4}\mathcal{F}^{a}_{ \mu \nu} \mathcal{F}^{ \mu \nu}_{a}   +\frac{\lambda+g \vartheta }{2}   (\nabla_{\mu} A_{a}^{ \mu} ) (\nabla_{\nu} A^{a \nu}) \right],
\label{eym_nl_action}
\end{equation}
where $g$ is the coupling constant introduced above, $\vartheta$ is a Lagrange multiplier that enforces the constraints $\nabla_{\mu} A^{\mu}_a = 0$, and we have introduced the object 
\begin{equation}
    \mathcal{F}^{a}_{\mu \nu} = 2 \nabla_{[\mu} A^a_{\nu]} + g  f^{bca}A_{b\mu} A_{c\nu}.
\end{equation}

This resulting theory has the properties that we were looking for. Namely, its $g \rightarrow 0$ limit is the free emergent linearized Yang-Mills theory discussed above. The same comment applies to the infinitesimal emergent non-linear gauge transformations, given by
\begin{equation}
    A^{a}_{ \mu} \rightarrow A^{a}_{ \mu} + \nabla_{\mu} \chi^a + g f^{abc} A_{b \mu} \chi_c
\end{equation}
with the fields $\chi^a$ obeying
\begin{equation}
    \Box \chi^a + g f^{abc}   A^{\mu}_{b} \nabla_{\mu} \chi_c =0.
\end{equation}
These reduce to the emergent gauge symmetries of the linear theory from Eq.~\eqref{gauge_transf_ym1} and Eq.~\eqref{gauge_transf_ym2} in the $g \rightarrow 0$. These transformations are such that they preserve the subspace $\mathcal{U}$ defined by the constraint $\nabla_{\mu} A^{a \mu} = 0$. Finally, we notice that the inclusion of matter is straightforward, since the bootstrapping procedure in the matter sector is disentangled from the bootstrapping procedure in the gauge sector that we have sketched here. The self-consistency constraint that appears in the matter sector is that the fields need to transform under a representation of the gauge group whose associated Lie Algebra has the structure constants $f^{a b c}$ that we introduced above. For scalar fields, this procedure needs two iterations, like for the Yang-Mills case, whereas for fermionic fields it requires just one iteration to be completed.

This has been just a brief summary of how an emergent Yang-Mills theory can be obtained by a suitable combination of the mechanism introduced in Sec.~\ref{Mechanism} and the bootstrapping procedure. For further details, see~\cite{Barcelo2021}. In the following we will focus on a situation that has not been studied previously.

\subsection{Linear graviton physics}
\label{Example2}

The discussion of the case for a symmetric tensor field $h^{\mu \nu}$ is completely parallel to the above section. The first step is to write down the most general quadratic Lorentz-invariant action for such a field. This means that the $h^{\mu \nu}$-field is a good variable to describe our effective field theory. For instance it can represent the deformations of a given condensed-matter-system near a Fermi point, as we have explained in the introduction. Therefore the resulting effective theory does not need be gauge-invariant: this is definitely asking too much since everything we know about it at this stage is that it can be described by a field $h^{\mu \nu}$. Explicitly, this action has the form
\begin{align}
    & S = \int d^4 x \sqrt{ - \eta} \left[ - \frac{1}{4} \nabla_{\mu} h^{\alpha \beta} \nabla^{\mu} h_{\alpha \beta} + \frac{1}{2} \left( 1 + \xi_1\right) \nabla_{\mu} h^{\mu \alpha} \nabla_{\nu} h^{\nu}_{\alpha} \right. \nonumber \\
& \left.  - \frac{1}{2} \left(1 + \xi_2 \right) \nabla_{\mu} h^{\mu \nu} \nabla_{\nu} h + \frac{1}{4} \left( 1 + \xi_3 \right) \nabla_{\mu} h \nabla^{\mu} h - \frac{1}{4} m_1^2 h_{\mu \nu} h^{\mu \nu} - \frac{1}{4}m_2^2 h^2 \right],
\label{h_action}
\end{align}
where indices are raised and lowered with the flat spacetime metric $\eta_{\mu \nu}$, we have introduced the trace of the field $h^{\mu \nu}$ as $h = h^{\mu \nu} \eta_{\mu \nu}$, and we have introduced the dimensionless parameters $\xi_i$ and the mass parameters $m_i$. 

We can also include a coupling to matter by adding a term of the form $h^{\mu \nu} T_{\mu \nu}$ to the previous Lagrangian, with the properties of $T_{\mu \nu}$ being discussed in more detail below (for the moment we will just assume that it is a symmetric tensor). The equations of motion can be computed by performing a variational derivative, which results in
\begin{align}
& \frac{1}{2} \Box h_{\mu \nu} - (1+ \xi_1) \eta_{\beta ( \mu} \nabla_{\nu)} \nabla_{\alpha} h^{\alpha \beta} + \frac{1}{2} (1 + \xi_2) \left( \nabla_{\mu} \nabla_{\nu} h + \eta_{\mu \nu} \nabla^2 \cdot h \right) \nonumber \\
& - \frac{1}{2} ( 1 + \xi_3) \eta_{\mu \nu} \Box h - \frac{1}{2} m_1^2 h_{\mu \nu} - \frac{1}{2} m_2^2 \eta_{\mu \nu} h  =  T_{\mu \nu},
\label{h_equations}
\end{align}
where we have introduced the notation $\nabla^2 \cdot h = \nabla_{\mu} \nabla_{\nu} h^{\mu \nu}$ and we will keep using it throughout this section. 

Before moving forward, it is better to pause and discuss the values of the parameters $\xi_i$ and $m_i$ for which the system has a gauge symmetry from the beginning, in order to avoid those cases in the discussion of emergent gauge symmetries. We will follow the discussion of~\cite{Alvarez2006}. On the one hand, for $\xi_1 = m_1 =0$, the equations of motion are invariant under the group of linearized transverse diffemorphisms, whose infinitesimal transformations are generated by a divergenceless vector field $\chi^{\mu}$ as follows:
\begin{equation}
    h^{\mu \nu} \rightarrow h^{\mu \nu} + 2 \nabla^{( \mu} \chi^{\nu)}, \qquad \nabla_{\mu} \chi^{\mu} =0.
\end{equation}
In addition, for $m_2 = 0$, $\xi_2=\xi_3=0$ we can relax the condition on the divergence for the vector field $\nabla_{\mu} \chi^{\mu} =0$ and recover the full group of linearized diffeomorphisms, i.e. one recovers the Fierz-Pauli theory~\cite{Fierz1939}, i.e. transformations generated by arbitrary vector fields $\chi^{\mu}$ as
\begin{equation}
    h^{\mu \nu} \rightarrow h^{\mu \nu} + 2 \nabla^{( \mu} \chi^{\nu)}.
    \label{normal_diffs}
\end{equation}

We could have considered a set of transformations that is slightly more general than the ones introduced here, namely transformations generated also by a vector field $\chi^{\mu}$ as  
\begin{equation}
    h_{\mu \nu} \rightarrow h_{\mu \nu} + 2 \nabla_{( \mu} \chi_{\nu)} + k \eta^{\mu \nu} \nabla_{\alpha} \chi^{\alpha}
\label{moregeneraldiff}
\end{equation}
However, these transformations do not add anything new. To understand this, we first notice that when we put the masses equal to zero $m_1 = m_2 = 0$, a field redefinition of the form 
\begin{equation}
    h^{\mu \nu} \rightarrow h^{\mu \nu} + \lambda \eta^{\mu \nu} h,
    \label{field_redef}
\end{equation}
(with $\lambda \neq - 1/4$ in order for it to be invertible), leaves invariant the functional form of the Lagrangian at the expense of changing the parameters $\xi_2$ and $\xi_3$ as~\cite{Alvarez2006} 
\begin{align}
    & \xi_2 \rightarrow \xi_2 + 4 \lambda + \lambda (4 \xi_2 - 2), \\
    & \xi_3 \rightarrow \xi_3 + 2 \lambda \left( 2 + 4 \xi_3 - \xi_2 \right) + \lambda^2 \left( 16 \xi_3 - 8 \xi_2 + 6 \right).
\end{align}
Had we begun with the case $\xi_1 = \xi_2 = \xi_3 = 0$, for which the theory is invariant under the whole set of diffeomorphisms, we would have ended up with new parameters $\xi_2$ and $\xi_3$ 
\begin{align}
    & \xi_2 = 2 \lambda, \\
    & \xi_3 = 4 \lambda + 6 \lambda^2.
\end{align}
Furthermore, notice that within this new parametrization of field space, the theory is not invariant under the standard transformations of the form~\eqref{normal_diffs} but transformations of the form~\eqref{moregeneraldiff} with $k = 2 \lambda$. Thus, it is clear that the more general set of transformations of the form~\eqref{moregeneraldiff} is included in the set of transformations of the form of ordinary linear diffeomorphisms~\eqref{normal_diffs} upon, possibly, a field redefinition. Hence, we will focus only on the set of transformations~\eqref{normal_diffs} since the transformations~\eqref{moregeneraldiff} do not add anything new. 

On the other hand, if $m_1 = m_2 = 0$ and the parameters $\xi_i$ satisfy the relations $-2  + \xi_2 - 4 \xi_3 = 0$ and $1 - \xi_1 + 2 \xi_2 = 0$, we have the following Weyl transformations, generated by a scalar field $\phi$, as gauge transformations:
\begin{equation}
    h^{\mu \nu} \rightarrow h^{\mu \nu} + \frac{1}{2} \eta^{\mu \nu} \phi. 
\label{weyl_transf}
\end{equation}
This Weyl symmetry can be combined with transverse diffeomorphisms to build the so-called linearized WTDiff (Weyl Transverse-Diffeomorphism) invariant gravity by choosing $\xi_1 = 0$, which guarantees that transverse diffeomorphisms are gauge symmetries as explained above. This forces $\xi_2 =  -1/2$ and $\xi_3 = -5/8$.\footnote{This corresponds precisely to the case of considering the theory with $\xi_1 = 0$ and performing a putative field redefinition of the form~\eqref{field_redef} with the limiting value $\lambda = -1 /4$ for which the transformation becomes singular. Thus, it is not a true field redefinition since one can not recover the trace of $h$ from the transformed field~\cite{Alvarez2006}.}

Fierz-Pauli theory (with the whole linearized group of diffeomorphisms) and WTDiff (with linearized Weyl and transverse diffeomorphism transformations) are the two largest possible gauge symmetry groups for a tensor field $h^{\mu \nu}$~\cite{Alvarez2006}.

Once we have discussed the choices of parameters for which the theory displays gauge symmetries, let us go back to the discussion of emergent gauge symmetries. We will be assuming that the parameters $\xi_i$ are such that the theory is \emph{generic}, in the sense that the coefficients of the different terms that will appear from now in equations of motion, conserved currents, \emph{etc}. do not vanish. Also we will assume that they take generic values different from the ones that endow the action~\eqref{h_action} with gauge symmetries from the beginning. The discussion of special choices of the parameters has been moved to Appendix~\Ref{sec:appendix} since it just involves the analysis of different particular cases and does not affect the main point of this section. 

Following the discussion in Sec.~\ref{Mechanism}, the first step is to identify a set of physical symmetries in our system that can become potential gauge symmetries. The key observation is that the following transformations 
\begin{equation}
    h^{\mu \nu} \rightarrow h^{\mu \nu} + 2 \nabla^{ (\mu} \chi^{\nu)},
    \label{putative_diff}
\end{equation}
with the field $\chi_{\mu}$ obeying the following conditions,
\begin{align}
    & -\xi_1 \Box \nabla_{( \mu} \chi_{\nu)}
    +(\xi_2 - \xi_1) \nabla_\mu \nabla_\nu (\nabla^\alpha \chi_\alpha)
    \nonumber \\
    & + (\xi_2  -\xi_3) \eta_{\mu \nu} \Box \nabla_{\alpha} \chi^{\alpha} - m_1^2 \nabla_{( \mu} \chi_{\nu)} - m_2 ^2 \eta_{\mu \nu} \nabla_{\alpha} \chi^{\alpha} = 0.
    \label{complex_transformation}
\end{align}
are physical symmetries of the action defined in Eq.~\eqref{h_action}. This can be seen by computing their Noether currents. The previous expression can be written in a more illustrative form as
\begin{align}
  &\nabla_{(\mu} V_{\nu)} + \xi_4 \eta_{\mu\nu} (\nabla_\alpha V^\alpha) + \left( \frac{\xi_4 }{\xi_1}~ m_1^2 - m_2^2 \right)\eta_{\mu\nu} (\nabla_\alpha \chi^\alpha)=0,
    \nonumber \\
  &V_\nu= \Box \chi_\nu + \left(1-\frac{\xi_2}{ \xi_1} \right) \nabla_\nu (\nabla_\sigma \chi^\sigma) + \frac{m_1^2}{\xi_1} \chi_\nu,
    \label{complex_transformation2}
\end{align}
with $\xi_4=(\xi_3 -\xi_2)/(2\xi_1-\xi_2)$. As we are interested in physical symmetries that can become emergent gauge symmetries, let us concentrate on the massless case. Then we have 
\begin{align}
  &\nabla_{(\mu} V_{\nu)} + \xi_4 \eta_{\mu\nu} (\nabla_\alpha V^\alpha)=0,
    \nonumber \\
  &V_\nu= \Box \chi_\nu + \left(1-\frac{\xi_2}{ \xi_1} \right) \nabla_\nu (\nabla_\sigma \chi^\sigma).
    \label{complex_transformation3}
\end{align}
Taking the trace of the equation on the first line (contracting with $\eta^{\mu \nu}$), we obtain that $\nabla^\alpha V_\alpha=0$ and thus $\nabla_{(\mu} V_{\nu)}=0$. Imposing also that $\chi^{\mu}$ goes to zero at infinity, we deduce that $V^{\mu}=0$. 

At this stage and as an aside, it is interesting to realize that the condition $V^{\mu} (\chi) = 0$ for a given vector field $\chi^{\mu}$ can be extracted by requiring that the infinitesimal transformations from Eq.~\eqref{putative_diff} leave invariant the subspace of $h^{\mu \nu}$ obeying the following (putative gauge) condition of the form 
\begin{equation}
\nabla^\mu h_{\mu\nu} -\frac{\xi_2}{2\xi_1} \nabla_\nu h=0.   
\label{putative-gauge}
\end{equation}

Let us now pass to explicitly analyze the structure of the Noether charges. Applying Noether's theorem to the symmetries satisfying Eq.~\eqref{putative_diff} and suitably rearranging the terms, we get the following expression for the current with $V_{\mu} =0$,
\begin{equation}
    J^{\mu} = W^{\mu} + 2 \nabla_{\nu} N^{[\mu \nu]} + M^{\mu} + D^{\mu},
\label{Eq:general_current}
\end{equation}
where the terms in the previous expressions read 
\begin{align}
W^{\mu} & = - \Box h^{\mu \nu} \chi_{\nu}  + 2 T^{\mu \nu} \chi_{\nu}  , \\
N^{\mu \nu} & = \nabla^{ \nu} h^{\mu \rho} \chi_{\rho} + h^{\nu \rho} \nabla^{\mu} \chi_{\rho}, \\
M^{\mu} & =- \bigg[ (1+\xi_2) \nabla_{\alpha} h^{\mu \alpha} + (1+\xi_3) \nabla^{\mu}h - (1+\xi_1)h^{\mu \alpha} \nabla_{\alpha}  \nonumber \\
& + \frac{1}{2} (1 + \xi_2) h \nabla^{\mu} + (1 + \xi_2) h^{\mu \nu} \nabla_{\nu} - (1 + \xi_3)h \nabla^{\mu} \bigg] \nabla_{\rho} \chi^{\rho} , \\
D^{\mu} & = - \nabla^{\alpha} h_{\alpha \beta} \nabla^{\mu} \chi^{\beta} + \nabla^{\mu} \nabla^{\alpha} h_{\alpha \beta} \chi^{\beta} + (1 + \xi_1) \nabla_{\beta}h^{\beta}_{\ \nu} \nabla^{\mu} \chi^{\nu} + (1 + \xi_1) \nabla_{\beta} h^{\beta}_{\ \nu} \nabla^{\nu} \chi^{\mu} \nonumber \\
& - \frac{1}{2} ( 1 + \xi_2) \nabla_{\nu} h \nabla^{\mu} \chi^{\nu} - \frac{1}{2} ( 1 + \xi_2) \nabla_{\nu} h \nabla^{\nu} \chi^{\mu} + (1 + \xi_2) h \Box \chi^{\mu} .
\end{align}

The charges associated with these currents $J^{\mu}$ do not identically vanish as long as we avoid the particular choices of parameters discussed above. We can also check that by restricting to a subspace of solutions satisfying the condition~\eqref{putative-gauge} one is still not able to make all Noether charges associated with the symmetry $V_\mu=0$ to vanish with full generality. 

However, we can focus our attention on a reduced set of symmetries, defined as transformations of the form~\eqref{putative_diff} with the generators obeying
\begin{align}
\Box \chi^\nu=0, \nonumber \\
\nabla_\nu (\nabla_\sigma \chi^\sigma)=0.
\label{restricted_transformation}
\end{align}

Notice that these conditions automatically ensure that $V^{\mu} (\chi) = 0$, since they correspond to independently putting to zero the two terms entering the definition of $V^{\mu}$ in~\eqref{complex_transformation2}. It is not difficult to realize that the corresponding Noether charges all vanish when restricted to a subspace of solutions of the theory, which we call $\mathcal{U}$, characterized by the conditions   
\begin{align}
\nabla_\mu h^{\mu\nu}=0,\qquad \nabla_\nu h=0.   
\label{restricted_putative_gauge}
\end{align}

The conditions from Eq.~\eqref{restricted_transformation} for the transformations~\eqref{putative_diff} introduced above precisely correspond to the subset of such transformations that preserve the conditions from Eq.~\eqref{restricted_putative_gauge}. We will come back to this point shortly. Furthermore, notice that the conditions defining this subspace are Lorentz-invariant themselves.

To see why when we restrict ourselves to the subspace of solutions $\mathcal{U}$ the Noether charges of the symmetries satisfying Eq.~\eqref{restricted_transformation} identically vanish, it is useful to notice that we have rearranged the conserved current~\eqref{Eq:general_current} in a sum of four pieces, where:

\begin{enumerate}
\item
the first term, $W^{\mu}$, vanishes on shell within the subspace $\mathcal{U}$;
\item
the second term, given by the divergence of $N^{\mu \nu}$, is a superpotential;
\item 
the third term, $M^\mu$, vanishes identically if we take the transformations such that $ \nabla_{\alpha} \left( \nabla_{\mu} \chi^{\mu} \right)= 0$;
\item 
the fourth term, $D^{\mu}$, is such that it also identically vanishes within the subspace $\mathcal{U}$. 
\end{enumerate}

This discussion allows us to conclude that if we define a projection onto the subspace $\mathcal{U}$, defined by the constraints $\nabla_\mu h = 0$ and $\nabla_{\mu} h^{\mu \nu}=0$, of the massless theory, these physical symmetries become emergent gauge symmetries within that subspace. We will discuss in a moment how this subspace could be selected dynamically. 

In addition to the diffeomorphism-like transformations from Eq.~\eqref{putative_diff} already discussed, we can also ask whether the Weyl transformations introduced in Eq.~\eqref{weyl_transf} can become emergent gauge symmetries using the same mechanism. The first step is identifying the following Weyl-like physical symmetries of the theory, given by the transformations $h_{\mu \nu} \rightarrow h_{\mu \nu} +  \eta_{\mu \nu}\phi /2$ where the scalar field $\phi$ needs to obey the following equation:
\begin{equation}
\left[ \left( - \frac{1}{2} + \frac{1}{4} \xi_2 - \xi_3 \right) \Box - \frac{1}{4} m_1^2  - m_2^2 \right] \eta_{\mu \nu} \phi  +
\left[ \frac{1}{2} + \xi_2 - \frac{1}{2} \xi_1 \right] \nabla_{\mu} \nabla_{\nu} \phi = 0.
\label{weyl_gen_constraint}
\end{equation}

Taking the trace of this equation and considering the case $m_1 = m_2 = 0$, we find the condition 
\begin{equation}
    \Box \phi = 0, 
\end{equation}
and plugging it back in~\eqref{weyl_gen_constraint} we find the condition 
\begin{equation}
    \nabla_{\mu} \nabla_{\nu} \phi = 0, 
\end{equation}
for the Weyl transformations to be symmetries of the theory. The solutions to this equations are of the following form 
\begin{equation}
    \phi (x) = A + B_{\mu} x^{\mu} \qquad A,B_{\mu} \in \mathbb{R}.
\end{equation}

Now, we can compute the Noether currents associated with these symmetries which read
\begin{equation}
J^{\mu} = \frac{1}{4} \left( 2 - \xi_2 + 4 \xi_3 \right) \left(h \nabla^{\mu} \phi - \phi \nabla^{\mu} h \right),
\end{equation}
where we recall that we have set the masses to zero $m_1 = m_2 = 0$. To ensure that these transformations correspond to gauge symmetries, we need to ensure that they have trivial Noether charges. For that purpose, we have two options.

\emph{First case: $B_{\mu} \neq 0$}. If we take $B_{\mu} \neq 0$ we need to restrict ourselves to the subspace $h = 0$ in order to have a trivial current. However, notice that these transformations do not preserve such subspace and hence we would be in the first situation described in Sec.~\Ref{Mechanism}, namely the non-emergence of gauge symmetry. 

\emph{Second case: $B_{\mu} = 0$}. In this second case we can relax the condition on $h$ to be $\nabla_{\mu} h = 0$ in order to have a trivial current. Notice that these transformations preserve the subspace $\nabla_{\mu } h = 0$ and hence this second option corresponds to the second situation described in Sec~\Ref{Mechanism}, i.e. the emergence of gauge symmetry. More explicitly, to have WTDiff emergent transformations, we need to restrict ourselves to the same subspace $\mathcal{U}$ introduced above. We recall that such space is defined by the conditions~\eqref{restricted_putative_gauge} and that we need to choose it because the emergent Weyl gauge symmetries preserve them and the Noether currents restricted to such subspace give identically vanishing charges. Hence, we conclude that it is possible to also find emergent Weyl symmetries with the mechanism introduced in Sec.~\Ref{Mechanism}. 

It is interesting to realize that the space $\mathcal{U}$ in which we find the emergence of gauge symmetries is the same independently of whether we are trying to find the closest theory to Fierz-Pauli theory or to linearized WTDiff within this emergent paradigm. In fact, a (non-transverse) diffeomorphism generated by a vector field with constant divergence $\nabla_\alpha \chi^\alpha= \kappa \in \mathbb{R} $, as the ones we have considered above, is equivalent to the composition of a transverse diffeomorphisms generated by the transverse part of such vector field $\chi^{\mu}$ and a Weyl transformation generated by a constant scalar field with $\phi = 4 \kappa $. Both of them can be seen to produce constants shifts in $h$ and hence preserve the subspace $\mathcal{U}$. In other words, when we understand the conditions defining subspace $\mathcal{U}$, i.e. $\nabla_{\mu} h^{\mu \nu} = \nabla^{\nu} h =0 $, as gauge fixing conditions for either Fierz-Pauli or linearized WTDiff, the residual gauge symmetries that leave such subspace invariant for both theories are exactly the same. Thus, whether one decides to understand the emergent gauge theory we have built as having a Fierz-Pauli flavor or a linearized WTDiff flavor is inconsequential for the discussion at the linear level.

Now, let us pass to discuss under which conditions the space $\mathcal{U}$ is selected naturally by the dynamics of the theory. If we assume that the tensor $T^{\mu \nu}$ is conserved, taking the divergence of the equations of motion uncovers a structural constraint: 
\begin{equation}
- \frac{1}{2} \xi_1 \Box \nabla_{\mu} h^{\mu \nu} + \frac{1}{2} (\xi_2 - \xi_1) \nabla^{\nu} \nabla^2 \cdot h + \frac{1}{2}(\xi_2 - \xi_3) \nabla^{\nu} \Box h  =  \nabla_{\mu} T^{\mu \nu}=0.
\label{firstdiv}
\end{equation}

Taking an additional divergence we find 
\begin{equation}
\Box \left[
(\xi_2 - 2 \xi_1) (\nabla^2 \cdot h) + 
(\xi_2 - \xi_3) \Box h
\right]=0.
\label{seconddiv} 
\end{equation}

Given this structural equation, which is a wave equation without a source, we can argue as we have already done before that the following solutions are selected dynamically,
\begin{equation}
(\xi_2 - 2 \xi_1) (\nabla^2 \cdot h) + (\xi_2 - \xi_3) \Box h =0. 
\label{first_natural_condition} 
\end{equation}
Plugging this condition in~Eq.\eqref{firstdiv} we obtain
\begin{equation}
\Box
\left[
\nabla_{\mu} h^{\mu \nu} + 
\frac{(\xi_2 -\xi_3)}{(\xi_2 -2\xi_1)} \nabla^\nu h
\right]=0,
\label{second_natural_condition}
\end{equation}
and applying again the argument that the lack of sources selects naturally the trivial solution, we conclude that
\begin{equation}
\nabla_{\mu} h^{\mu \nu} + \frac{(\xi_2 -\xi_3)}{(\xi_2 -2\xi_1)} \nabla^\nu h=0.
\label{third_natural_condition}
\end{equation}
Notice that this condition does not imply that the two terms in the previous expression are zero independently. For this to happen one needs to impose further constraints.

One possibility is to further require that the trace of $T^{\mu \nu}$ is a constant over spacetime. Taking the trace of the equations of motion for the field $h^{\mu \nu}$ and introducing the following notation $T = \eta ^{\mu \nu} T_{\mu \nu}$, we find 
\begin{equation}
\frac{1}{2}\left[ \left( - 2 +  \xi_2 - 4 \xi_3 \right) \Box h + \left( 2 - 2\xi_1 + 4 \xi_2 \right) \nabla^2 \cdot h \right] = T.
\label{trace}
\end{equation}
Taking a derivative, using the assumption of constant trace for the current, i.e. $\nabla_{\mu} T = 0$, and using also~\eqref{first_natural_condition} we obtain the following condition (assuming that the parameters do not take specific values, which we discuss in Appendix~\ref{sec:appendix}):
\begin{align}
\Box (\nabla_\nu h)=0.
\label{trace_clean}
\end{align}
Again, for a generic system, and applying the argument that the lack of sources leads naturally to the trivial solution, we obtain the condition $\nabla_\nu h=0$. Inserting this condition in Eq.~(\ref{third_natural_condition}) we precisely see that the subspace $\mathcal{U}$ introduced in Eq.~(\ref{restricted_putative_gauge}) is naturally recovered. Thus, under the assumption of coupling to a conserved current $\nabla_{\mu} T^{\mu \nu} = 0$ with constant trace, $\nabla_{\mu} T = 0$, we have found that the system develops emergent gauge symmetries that can either be understood as emergent diffeomorphisms or emergent Weyl-Transverese diffeomorphisms.

Let us summarize the discussion concerning the emergence of gauge symmetries for the two-index symmetric tensor field $h^{\mu \nu}$ up to this point. We have managed to prove that for the most general Lorentz-invariant theory constructed from such tensor field, the mechanism presented in Sec.~\ref{Mechanism} works and provides us with emergent gauge symmetries. At the linear level, the gauge transformations that emerge can be either understood as the manifestation within this emergent framework of either linearized diffeomorphisms or linearized Weyl and transverse diffeomorphisms transformations. To put it explicitly, consider any generic Lorentz-invariant theory of $h^{\mu\nu}$ which fulfills the following conditions:
\begin{enumerate}
    \item It describes massless excitations, namely the masses from the Lagrangian in Eq.~\eqref{h_action} are equal to zero ($m_1=m_2 = 0$). 
    \item It couples to a two-index symmetric source $T^{\mu \nu}$ that is conserved, i.e. divergenceless $\nabla_{\mu} T^{\mu \nu} = 0$. 
    \item The source has constant trace, namely $\nabla_{\mu} T = 0$. 
\end{enumerate}

Our analysis implies that at the linear level, such a theory has a natural truncation that is indistinguishable from either Fierz-Pauli theory or linearized WTDiff theory in the gauge described by~\eqref{restricted_putative_gauge}. Notice that, instead of a projection to a given dynamical subspace of the theory, for these theories we understand the conditions $\nabla_{\mu} h^{\mu \nu} = \nabla^{\nu} h= 0$ as gauge fixing conditions. Actually, we emphasize that Fierz-Pauli theory and linearized WTDiff are indistinguishable in this specific gauge, in the sense that the residual gauge transformations that emerge are of the same form. This is the reason for understanding this emergent gauge theory as the manifestation of either WTDiff and Fierz-Pauli theory in the emergent framework: both theories are indistinguishable in this specific gauge. 

The extension to the non-linear regime of the gravitational case through a bootstrapping procedure is expected to be much more involved than for the vector-field case discussed in Subsec.~\ref{Example1} and it is left for future work. It constitutes an ongoing project on which we expect to report soon. Here we simply advance that the source of this difficulty arises, apart from the obvious technical complications appearing due to the presence of much more terms in the Lagrangian, from the nature of the non-linear terms appearing in the series generated through the bootstrapping procedure. Whereas for vector fields there appears a finite series that finishes after two iterations, for gravity we need to deal with an infinite ``formal'' series to obtain the non-linear theory~\cite{Barcelo2014,Barcelo:2014qva}. 

\section{No-go theorems for emergent gravity reconsidered}
\label{No-gos}

The Weinberg-Witten theorem~\cite{Weinberg1980} is often used to make a case against emergent gravity, attempting to claim that any research program pursuing to describe gravitational dynamics as an emergent phenomenon is condemned to fail. Marolf's recent arguments seem to present also an important obstruction towards the emergent gravity paradigm~\cite{Marolf2015}, at least from the perspective of constraining the required characteristics of emergent approaches. In this section we will  review the  Weinberg-Witten theorem and also revisit Marolf's argument and point out why the possibility that gauge symmetries may be emergent cannot be ignored when attempting to use these theorems in this way. Although our results for gravity are still for the linear theory, there is a clear interplay between our results and these obstructions as we will discuss. Let us note that  these theorems assume a quantum character for the gravitational field. Our results for linear fields, although derived for classical fields, can be straightforwardly extended to the quantum level through a quantization procedure, see section 3.4 of Ref.~\cite{Barcelo2016}. The issue of the non-linear theory is much more subtle, since we do not have a non-perturbative definition of quantum field theories that does not involve a lattice regularization. Hence, a careful analysis of that point would be required, although our main conclusions below do not depend crucially on this issue.

\subsection{Weinberg-Witten theorem}
\label{WW}

The Weinberg-Witten theorem~\cite{Weinberg1980} states that a theory that allows the construction of a conserved Lorentz-covariant energy-momentum tensor cannot contain gravitons, neither fundamental nor composite in the spectrum of asymptotic states (in fact, the theorem applies to particles with spin $j>1$). The content of the theorem can be rephrased as saying that gravitons are so special that one cannot build a Lorentz-covariant and gauge-invariant energy-momentum tensor for them. This distinctive feature already appears in the Fierz-Pauli theory for linear gravitons. One can certainly build different Lorentz-invariant energy-momentum tensors for $h_{\mu\nu}$; however,  none of them  happens to be invariant under the (linear) gauge transformation generated by a vector field $\chi^{\mu }$ that take the form $\delta h^{\mu\nu}=2 \nabla^{(\mu} \chi^{\nu)}$, with $\nabla$ representing the connection compatible with flat spacetime metric, as before.

From the perspective of emergent gravity the theorem implies that, if excitations with the properties of gravitons arise in a suitable regime, these excitations cannot have a Lorentz-invariant energy-momentum tensor that is gauge invariant. This is often taken as an indication that obtaining gravitons from systems akin to condensed-matter systems is not possible. However, our discussion of emergent gauge symmetries above indicates that such an interpretation is pulling too far, or misunderstanding, the significance of the theorem. Let us discuss this more explicitly.

The key aspect is the emergent nature of gauge symmetries and what this implies for observables, in particular for the energy-momentum tensor. Our starting point is a system akin to a condensed-matter system which develops excitations with the properties of gravitons at low energies. We have seen that this requires certain conditions on the source to which the field $h^{\mu\nu}$ couples which, when satisfied, are equivalent to the decoupling of some degrees of freedom leading to the natural restriction to the subspace $\mathcal U$ described in previous sections. As the original high-energy theory has no gauge symmetries, all configurations are different and, therefore, in general will have different values for most observables (in this section, this word will always refer to the high-energy theory), which includes the fields themselves and the energy-momentum tensor as specific examples. However, the development of emergent gauge symmetries implies the emergence of equivalence classes for which the values of certain observables do not change, which are then denoted as gauge invariant. The specific observables that change or not within a given equivalence class are determined by the structure of the theory. The basic fields $A^\mu_a$ and $h^{\mu\nu}$ are examples of observables of the high-energy theory that   change under the emergent gauge symmetries. On the other hand, for vector fields $A^\mu_a$ the energy-momentum tensor is gauge-invariant \cite{Barcelo2016,Barcelo2021}, while for gravity it is not.

In summary, that certain observables of the high-energy theory turn out  to be non gauge invariant does not pose a problem for emergence. From the perspective of the high-energy theory, all observables retain their meaning and have well-defined values regardless of the emergence of gauge symmetries. It is certainly not required and even unreasonable to assume that all observables of the high-energy theory become gauge invariant. In general, the process of emergence separates observables in two categories, depending on their behavior under gauge transformations. But whether the set of gauge-invariant observables includes the energy-momentum tensor, or any other observable for that matter, is inconsequential for the rationale of emergence.

We have discussed the Weinberg-Witten theorem from a purely conceptual point of view and highlighted that, although it points out a particular property of gravity, it does not forbid the emergence of gravity. From a technical point of view, it applies to gravity when understood as a perturbative quantum field theory built on top of flat spacetime having a well-defined notion of gravitons as asymptotic states. Let us also mention that, within this framework, several technical aspects of the theorem have been criticized throughout the years~\cite{Sudarshan1981,Kugo1982,Flato1983,Lopuszanski1984,Lopuzanski1988}. These criticisms further limit the application of the theorem to realistic physical situations. Also a comprehensive and systematic discussion of the ways in which Weinberg-Witten theorem can be avoided can be found in~\cite{Bekaert2010}.

\subsection{Marolf's theorem}
\label{Marolf}

The very interesting work of Marolf~\cite{Marolf2015} contains a theorem that seems to constrain the kind of theories that might give rise to an emergent gravitational description. According to Marolf~\cite{Marolf2015}, the only theories that might give rise to such a description are the ones that he calls \emph{kinematically non-local} (we will introduce his definition of kinematical non-locality in a moment). From our perspective, although Marolf's theorem pinpoints the particularities of constructing an emergent theory of gravity from a \emph{kinematically local} theory, it does not automatically imply that such theories cannot have an effective gravitational description in some regime. Let us now introduce Marolf's theorem and carefully analyze it as well as its implications. Quoting it literatelly, Marolf's theorem states~\cite{Marolf2015}: 

\vspace{0.5cm}

\noindent
\textbf{Marolf's theorem:} \emph{``Consider any limit where the effective description of a local theory is a gravitational theory with universal coupling to energy, the same notion of time evolution, and a compatible definition of locality. In this limit all local observables away from the boundary become independent of time.''}

\vspace{0.5cm}

The theorem above contains several definitions that need to be clarified to make it intelligible, which we review here for the benefit of the reader. On the one hand, Marolf restricts his study to what he calls gravitational theories with \emph{universal coupling to the energy}, namely theories for which the Hamiltonian can be written as the integral over the boundary of space at each time of some local function, which he calls gravitational flux. Furthermore, such gravitational flux needs to be a gauge-invariant function of the gravitational field and its derivatives. All diffeomorphism-invariant theories fall into this category (non-Lorentz-invariant theories like for instance Ho\v rava-Lifshitz gravity~\cite{Horava2009} do not in a strict sense, although Marolf suggests the existence of a straightforward generalization of this theorem that would apply to such theories). Massive gravity theories, like dRGT theory~\cite{deRham2010}, and scalar gravities, like the so-called Nordstr\"om gravity~\cite{Deruelle2011}, do not fall into this category since they do not have universal coupling to the energy.

On the other hand, by locality he assumes a version of microcausality which he calls \emph{kinematic locality}. According to his definition, a kinematically local theory is one for which every pair of local bosonic gauge-invariant Heisenberg operators evaluated at spacelike points commute. 

Then, he states that if there were a kinematically local (non-gravitational) theory that in some regime gave place to an effective description in terms of a gravitational theory with universal coupling to energy, all of its local bulk observables would freeze out unless there were a change on the notion of time evolution or locality between the gravitational and non-gravitational descriptions (these are notes of caution). Technically speaking, Marolf's observation is that given the gravitational flux which is a gauge-invariant operator, it should be possible to write it in the microscopic underlying theory as an integral over gauge-invariant operators supported near the boundary. Otherwise, it would correspond to a theory in which the notions of locality between the microscopic and effective gravitational theory would be very different. Assuming this \emph{compatible-locality property} among both descriptions, Marolf reaches the conclusion that, given a kinematically local underlying theory and having expressed the Hamiltonian as an operator in that theory just supported near the boundary, all the local (or bulk) observables of the theory need to freeze out in the regime in which the effective gravitational description applies. This is because the Hamiltonian (being itself another observable) will commute with any local observable away from the boundary.

Given the theorem, he then concludes that it is difficult to obtain an emergent gravitational description starting from microscopic kinematically local theories, without changing abruptly the notion of locality. Otherwise, the emergence would entail a rather strange behaviour on the non-gravitational description. Thus, he interprets the theorem as saying that one would rather start looking for emergence from microscopic theories which are already kinematically non-local at a fundamental level, as it is done for instance in the context of gauge-gravity duality~\cite{Banks1996,Maldacena1997} which, to some extent, can be interpreted as emergent theories for the bulk gravity theory. However, this conclusion is based on the additional assumption that all observables in the UV completion of the theory must freeze at low energies or, equivalently, that all these observables must be gauge invariant, which may not be the case in an emergent framework. If at least one observable of the UV completion is not gauge invariant, then the notion of locality can be preserved. This would be generally the case if gauge symmetries are emergent.

If we take our discussion of the linear setting above and assume the existence of a suitable non-linear extension along the lines realized in the discussed Yang-Mills case, we can make the following observation. When an effective gauge invariance is developed in the low-energy regime of the theory, the observables of the high-energy theory can be separated into those that are gauge invariant from the perspective of the emergent gauge symmetry and those that are not. As an intermediate stage of emergence, we started with a generic massless Lorentz-covariant system described by the action \eqref{h_action} for an object $h^{\mu\nu}$. Without further considerations, this system does not describe gravity, as for instance it contains additional degrees of freedom, and, therefore, does not satisfy the conditions of Marolf's theorem. However, we have discussed that, if the field $h^{\mu\nu}$ couples to a conserved and constant-trace two-index symmetric current, there is a natural decoupling of certain degrees of freedom and a simultaneous emergence of gauge symmetries. Interestingly, this decoupling makes the construction compatible with Marolf's conditions for the theorem. Then, it is true that all the local gauge-invariant observables away from the boundary need to freeze out. However, non-gauge-invariant observables, which are still meaningful from the point of view of the high-energy UV-completion of the theory, do not need to freeze out. In fact, our example shows that actually they
(i.e. the $h_{\mu\nu}$)  do not freeze out. As we have pointed out above, these observables are not constrained to be frozen by the theorem. The freeze out only affects a subset of the observables of the high-energy theory, namely those observables that are invariant under the emergent gauge transformations. In this way, the high-energy theory retains an ``interesting enough'' bulk dynamics.

Another aspect that we think is worth keeping in mind is the effect of having a background metric. As our underlying theory has a (flat) background metric the degrees of freedom are defined with respect to  this background and are perfectly local, as for electrodynamics and other quantum field theories. In the same vein, our theory has a well-defined local Hamiltonian density and does not have a fundamental gauge invariance. However, one could decide---and this is a choice---to analyze the theory from a gauge perspective, arguing that the background metric is not operationally measurable. Then, for consistency one has to change the notion of degree of freedom as now one cannot use the background metric as an external reference. The degrees of freedom must now be  gauge-invariant variables defined with respect to the metric, variables which in general relativity are highly non-local. Now, the former Hamiltonian density is not gauge invariant and therefore one can no longer think of it as locally defining an energy density. Instead, from the gauge perspective the Hamiltonian appears as a gauge-invariant and non-local (holographic) quantity. The curvatures in the bulk of spacetime certainly evolve, but even those do not constitute gauge-invariant degrees of freedom as we do not have an independent notion of what is a point (or event). If the notion of what is a degree of freedom did not change from the underlying theory to the gravitational emergent theory, then Marolf's theorem would be true in asserting that any local (gauge-invariant!) observable would need to freeze in the effective gravitational regime. However, the change in the notion of degree of freedom  permits that local degrees of freedom in the bulk of the underlying theory have a well-defined local kinematics.

Therefore, a decoupling from the high-energy perspective has as consequence a freeze out of certain non-local observables. While we have not provided a complete non-linear model satisfying the criteria that guarantee the decoupling, the latter are general enough to make us think that such models should exist (assuming that the extension of our theory into the non-linear regime is completely parallel to the one we have already done with Yang-Mills theory~\cite{Barcelo2021}). At this stage we can safely conclude that searching for models satisfying these criteria is a well-motivated endeavour.

\section{Discussion and conclusions} 
\label{Conclusions}

In this work we have reviewed a mechanism proposed by some of the authors for the emergence of gauge symmetries in field theories based on the natural decoupling of some of their degrees of freedom. We have also reviewed how this mechanism for the emergence of linear gauge theories combined with a suitable bootstrapping procedure linking the linear and non-linear regimes can give rise to emergent gauge symmetries resembling Yang-Mills theories. This example serves as a warm-up exercise for the more convoluted case of finding an emergent gravitational theory. 

We have analyzed in detail the case of linear gravitons and found that it is possible to give rise to the gauge symmetries characteristic of linear gravity through this mechanism. Although we have left the analysis of the non-linear extensions through a bootstrapping procedure for future work, we expect no substantial differences from the standard bootstrapping for gravity. The main difficulties actually come from the problems that are present already for Fierz-Pauli theory: namely the infinite nature of the series, the inherent ambiguities present in the construction, etc. 

Additionally, we have discussed some results in the literature that are often invoked as no-go theorems and discussed their interplay with our work. These results are indeed technically correct and pinpoint some differential features of gravity and gravitons with respect to other theories and gauge particles. However, using these results as no-go theorems implies making additional assumptions about the possible UV-completions, in particular regarding the preservation of gauge symmetries at high energies. As we have discussed, these theorems cannot rule out emergent frameworks in which gauge symmetries are emergent themselves.

We finish this work by mentioning that, although many technical details remain to be understood and analyzed within this framework, it seems to provide a successful approach towards building an emergent gravitational description of a certain theory. Given that a theory acquires an effective (typically we think of it as a low-energy regime) Lorentz-invariant description for which a field $h^{\mu \nu}$ describes adequately the excitations of the theory, we have found that the decoupling of some of its degrees of freedom (which is quite natural from a dynamical point of view) automatically gives rise to an emergent gauge symmetry. The remaining extension into the non-linear regime through a bootstrapping procedure would allow us to constrain the possible non-linear completions of such linear theory. As a result, we would be able to build emergent gravitational description of systems, by simply developing a low-energy Lorentz-invariant regime and having a $h^{\mu \nu}$ field in their low-energy spectrum. We expect to report soon on these issues. 

Another future line of research that is worth exploring is understanding the emergence of gauge symmetries in alternative descriptions of gravity. For instance, the reformulation of gravity as a more standard gauge theory of the local Poincaré group~\cite{MacDowell1977} seems to be a promising place to test the generality of the applicability of our mechanism.

\acknowledgments{Financial support was provided by the Spanish Government through the projects FIS2017-86497-C2-1-P, FIS2017-86497-C2-2-P, PID2019-107847RB-C44, PID2020-118159GB-C43, PID2020-118159GB-C44, and by the Junta de Andaluc\'{\i}a through the project FQM219. C.B. and G.G.M. acknowledges financial support from the State Agency for Research of the Spanish MCIU through the ``Center of Excellence Severo Ochoa'' award to the Instituto de Astrof\'{\i}sica de Andaluc\'{\i}a (SEV-2017-0709). GGM acknowledges financial support from IPARCOS.} 

\newpage

\appendix 

\section{Special cases of parameters for linear graviton fields}
\label{sec:appendix}

We have omitted the discussion of special cases of the parameters $(\xi_1,\xi_2,\xi_3)$ for which some of the terms in the equations from Sec.~\Ref{Example2} vanish. The unique special case that gives new insights is the case in which the three of them are equal: $\xi = \xi_1 = \xi_2 = \xi_3$. For the remaining cases, for example the case for which some of the terms in Eq.~\eqref{trace} vanish, namely $-2 + \xi_2 - 4 \xi_3 = 0$ or $1 - \xi_1 + 2 \xi_2 = 0$, one can see that the same arguments presented Sec.~\Ref{Example2} for the selection of the subspace $\mathcal{U}$ straightforwardly apply and hence we do not need to discuss them separately. 

The case $\xi = \xi_1 = \xi_2 = \xi_3$ where the masses have been put to zero is interesting because the condition~\eqref{complex_transformation} that the generators $\chi^{\mu}$ of the transformations~\eqref{putative_diff} need to verify reduces to
\begin{equation}
    \nabla_{\mu} \Box \chi_{\nu} = 0.
\end{equation}
Thus, it seems that for this specific choice of parameters the transformations of the form 
\begin{align}
    h^{\mu \nu} \rightarrow h^{\mu \nu} + 2 \nabla^{(\mu} \chi^{\nu)}, \qquad \Box \chi^{\mu} = 0,
    \label{diffs_full_glory}
\end{align}
have a chance to become emergent gauge symmetries. Actually, for this choice of parameters the terms of the associated current can be recast in the form~\eqref{Eq:general_current} without the $M^{\prime \mu}$. This means that we have a current of the form 
\begin{equation}
    J^{\prime \mu}  = W^{\prime \mu} + 2 \nabla_{\mu} N^{ \prime \mu \nu} + D^{\prime \mu}, \\
\end{equation}
with $W^{\prime \mu}$ a term that vanishes on shell, $ N^{\prime \nu \mu}$ a superpotential and $D^{\prime \mu}$ a term that identically vanishes within the subspace $\nabla_{\mu} h^{\mu \nu} = 0$. Their explicit form is:
\begin{align}
& W^{\prime \mu}  = - \Box h^{\mu \alpha} \chi_{\alpha} - (1 + \xi) \nabla_{\nu} \nabla^{\mu} h \chi^{\nu} + (1 + \xi) \Box h \chi^{\mu} + 2 T^{\mu \nu} \chi_{\nu}, \\
& N^{ \prime \mu \nu}  = \nabla^{\nu} h^{\mu \rho} \chi_{\rho} + h^{\nu \rho} \nabla^{\mu} \chi_{\rho} + \frac{1}{2} ( 1 + \xi) h \nabla^{\nu} \chi^{\mu} + (1 + \xi) \nabla^{\mu} h \chi^{\nu}, \\
& D^{\prime \mu}  = - \nabla^{\alpha} h_{\alpha \beta} \nabla^{\mu} \chi^{\beta} + \nabla^{\mu} \nabla^{\alpha} h_{\alpha \beta} \chi^{\beta} + (1 + \xi) \nabla_{\beta}h^{\beta}_{\ \nu} \nabla^{\mu} \chi^{\nu} + (1 + \xi) \nabla_{\beta} h^{\beta}_{\ \nu} \nabla^{\nu} \chi^{\mu}.
\end{align}

Thus, if we define a projection onto the subspace $\tilde{\mathcal{U}}$ defined by the condition
\begin{equation}
    \nabla_{\mu} h^{\mu \nu} = 0,
\end{equation}
the transformations of the form~\eqref{diffs_full_glory} become emergent gauge symmetries. 

The naturalness of such construction relies again on how natural it is to restrict the dynamics to the subspace $ \tilde{\mathcal{U}}$. We want to see if that subspace for this particular choice of parameters is selected dynamically. For this particular choice of parameters, the divergence of the equations of motion assuming we couple the theory to a conserved source automatically gives 
\begin{equation}
    \Box \nabla_{\mu} h^{\mu \nu} = 0,
\end{equation}
as it can be seen by putting all the $\xi_i$ parameters equal to the same value $\xi$ in Eq.~\eqref{firstdiv}. Thus, for this choice of parameters, the transformations of the form~\eqref{diffs_full_glory} naturally emerge as gauge symmetries since the subspace onto which we need to project for them to become emergent gauge symmetries is naturally selected from a dynamical point of view. 

Although this construction seems to be closer to Fierz-Pauli theory since it is equivalent to Fierz-Pauli theory in the partial gauge fixing given by the only condition $\nabla_{\mu} h^{\mu \nu} = 0$, it does not seem to be natural due to the fine-tuning among the parameters that it requires. For instance, we expect that, at the quantum level, radiative corrections may tend to rapidly break the equality among parameters. However, it is possible that the emergent gauge symmetries themselves are enough to protect this equality among the parameters and render them radiatively stable. We leave this question for future work, when we extend this construction to the whole non-linear and quantum regime, in which these radiative corrections enter and may play a relevant role.

\newpage

\bibliography{emergent_theories}

\end{document}